\documentclass[runningheads]{llncs}
\usepackage[T1]{fontenc}
\usepackage{footmisc}
\usepackage{hyperref}
\hypersetup{colorlinks}
\usepackage{graphicx}
\begin{document}
\title{Wavelet Integrated Convolutional Neural Network for ECG Signal Denoising}
\author{ }
\author{Takamasa Terada \and Masahiro Toyoura\thanks{This preprint has not undergone peer review or any post-submission improvements or corrections. The Version of Record of this contribution is published in MMM 2025, LNCS 15523, pp. 311–324, 2025., and is available online at \url{https://doi.org/10.1007/978-981-96-2071-5_23}.}}
\authorrunning{T. Terada and M. Toyoura}
\institute{ }
\institute{University of Yamanashi, 4-3-11 Takeda, Kofu, Yamanashi, 400-8511, Japan 
\email{mtoyoura@yamanashi.ac.jp}}

\maketitle              
\begin{abstract}
Wearable electrocardiogram (ECG) measurement using dry electrodes has a problem with high-intensity noise distortion. Hence, a robust noise reduction method is required. However, overlapping frequency bands of ECG and noise make noise reduction difficult. Hence, it is necessary to provide a mechanism that changes the characteristics of the noise based on its intensity and type. This study proposes a convolutional neural network (CNN) model with an additional wavelet transform layer that extracts the specific frequency features in a clean ECG. Testing confirms that the proposed method effectively predicts accurate ECG behavior with reduced noise by accounting for all frequency domains. In an experiment, noisy signals in the signal-to-noise ratio (SNR) range of -10---10 are evaluated, demonstrating that the efficiency of the proposed method is higher when the SNR is small.

\end{abstract}
\section{Introduction}
\label{sec:introduction}
Studies on machine-learning-supported detection and recognition of electrocardiogram (ECG) abnormalities have been conducted to improve the early prognosis of heart disease and to make evaluation easier for clinicians. To train these neural networks (NNs), clean ECG data with little-to-no noise are required to improve identification accuracy. Hence, preprocessing noise is a vital step when preparing the model.

In recent years, wearable devices have been used to acquire ECG data\cite{nigusse2021wearable}. Although these devices make data acquisition easy, ECG noise caused by patient activity becomes a problem. For such devices, non-invasive dry electrodes are often used for comfort and convenience; however, they are more susceptible to noise. Hence, data preprocessing is unavoidable. This study provides a novel noise reduction method that effectively reduces high-intensity ECG noise to acceptable levels.

Conventional methods, such as finite input response (FIR) filters and wavelet and thresholding techniques, have been proposed to remove various types of noise in advance, according to the method. Deep-learning autoencoder-based models also have been demonstrated to remove noise with high accuracy. However, these and other methods continue to have problems removing noise when multiple frequencies overlap. Notably, the accuracy of the algorithm varies depending on the preset parameters. It has also been reported that the accuracy of feature extraction decreases with noise intensity\cite{mohd2020analysis}.

\begin{figure}[t]
	\centering
	\includegraphics[width=0.55\hsize]{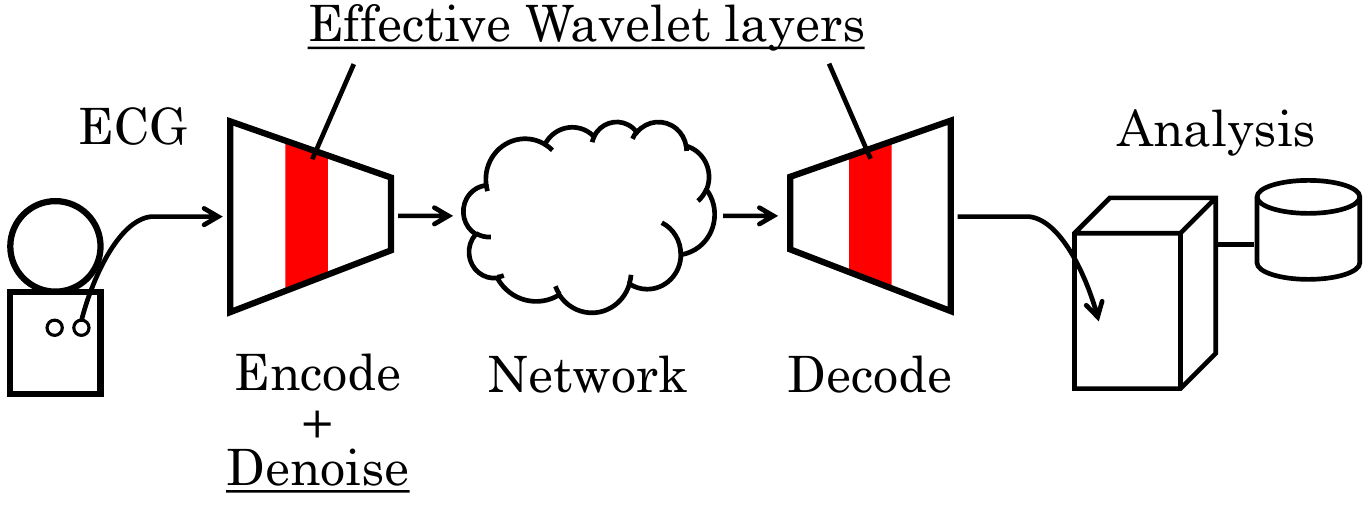}
	\caption{The proposed convolutional neural network (CNN) model with an additional wavelet transform layer in which the features are separated into high and low components.}\label{fig:Overview}
\end{figure}

With overlapping ECG noise frequencies, fixed-parameter filters tend to lose original ECG information, which degrades accuracy. To improve this, a mechanism is needed that changes the parameter characteristics according to noise intensity and type. Thus, this study proposes \textit{a convolutional neural network (CNN) model with an additional wavelet transform layer in which the features are separated into high and low components} 
as shown in Fig. \ref{fig:Overview}.  
Then, the CNN is trained using parameters based on specific frequency bands. The combination of high feature extraction separation by the wavelet transform makes it possible for the network to learn changing filter behaviors based on a clean ECG, even when the frequency bands overlap. In our experiment, noisy signals in a signal-to-noise ratio (SNR) range of -10---10 are evaluated, demonstrating that the efficiency of the proposed method is higher when high-intensity noise is corrupted.

The remainder of this paper is organized as follows: Section \ref{sec:relatedwork} provides an overview of the related research. Section \ref{sec:proposed} explains the proposed CNN model with an integrated wavelet transform layer. Section \ref{sec:experimental_condition} describes the experimental conditions, and Section \ref{sec:experimental_result} presents the results. Section \ref{sec:discussion} presents a discussion, and Section \ref{sec:conclusion} concludes the paper. 

\section{Related Work}
\label{sec:relatedwork}
In ECG measurements, frequency noise is known to be a problem. The types of noise can be divided into low- and high-frequency, and the low-frequency type includes baseline wandering (BW), which is caused by breathing and other movements. High-frequency noise includes muscle artifacts (MAs) during electromyography (EMG) and electrode motions (EMs) from electrode misalignments. There can also be commercial powerline interference (PLI) at specific frequencies of 50 and 60 Hz and thermal noise from electronics, which is normally treated as additive white Gaussian noise. To remove noise, two types of denoising algorithms have been proposed: machine learning (e.g., deep learning) and others (e.g., filtering and wavelet transform). Notably, there is a trade-off between computational accuracy and computational complexity.

The computational cost of denoising without machine learning is low, and it is possible to remove noise in real time. FIR filters are typically used, and low-pass filters (LPFs) and high-pass filters (HPFs) are widely employed to remove high- and low-frequency noise, respectively. 

Jenkal et al. \cite{jenkal2016efficient} proposed a method that combines a wavelet transform with an adaptive dual-threshold filter. Focusing on the fact that different types of noise are mixed in different frequency bands, high-frequency signals are removed by wavelet transform in advance, and an adaptive dual-threshold filter is then applied. This combination of techniques successfully removes high-frequency noise, including that of EMG, PLI, and BW. 
Prashar et al. \cite{prashar2021design} proposed a method that applies a dual-tree complex wavelet transform to produce a noise-robust method with a threshold determined by eight different rules. Other wavelet transform and thresholding methods are used, but the parameters are determined in different ways (e.g., S-median \cite{poornachandra2008wavelet} and improved thresholding \cite{reddy2009ecg}).

When predetermined parameters are used, the types of noise that can be removed are limited. In other cases, if the intensity of the unintended noise is large, it may not be sufficiently removed. If the noise is complex and has a large intensity, a more powerful method is needed.

In recent years, deep-learning-based methods have been proposed to automatically extract features. For example, Xiong et al. \cite{xiong2016ecg} proposed a method that combines wavelet transform and an NN. In this method, a deep-learning model is trained on data that has been denoised by wavelet transform and thresholding. By performing the wavelet transform in advance, the denoising effect is higher than when training the ECG with noise. On the other hand, Birok et al. \cite{birok2021ecg} proposed a method that combines EMD and NN, with which ECG signals are first converted to clean and noisy IMFs. Then, the noisy IMFs are denoised by an NN. As a result, denoising performance is improved over the EMD and NN methods, separately.

CNNs are known to improve denoising performance better than NNs that use fully connected layers. CNNs require fewer parameters to be trained than do full-connected models. Thus, they can be used to build lightweight models. Yildirim et al. \cite{yildirim2018efficient} proposed an autoencoder that extracts features with convolutional layers and pooling functions. Chiang et al. \cite{chiang2019noise} showed that ECG denoising performance could be improved by giving the convolutional filter a stride of two. The model without pooling performs better.

Wang et al. showed that generative adversarial networks (GANs) could also improve denoising performance \cite{wang2019adversarial}. The GAN has a generator that removes noise from the ECG and a discriminator that judges whether the ECG is ground truth or fake. This adversarial learning method greatly improves denoising performance by using a generator and a discriminator with all fully connected layers. On the other hand, Singh et al. \cite{singh2020new} proposed a GAN with fully convolutional layers, proving effective in denoising ECGs.

In recent years, edge-terminal processing (e.g., on wearable devices) has been considered. Hence, it is even more crucial to reduce data traffic by increasing the compression ratios. To do so, the encoder and decoder are placed at different locations, and the encoder sends compressed features to the decoder. For example, in the model discussed in \cite{yildirim2018efficient}, a 2,000-dimension vector was compressed to 62 dimensions by the encoders. Chiang et al. \cite{chiang2019noise} successfully compressed a 1,024-dimension vector to 32 dimensions. On the other hand, the encoder of GAN in \cite{singh2020new} expands 1,024-dimensional vectors to 8 $\times$ 1,024 dimensions, also using the encoder. Hence, a contraction path \cite{ronneberger2015u} is needed to concatenate the output of each encoder and the input of each decoder to improve performance. When using this model, data traffic increases in the contraction path and the output of the encoder, which then requires parameter reduction.

In another direction, Liu et al. \cite{Liu22} have proposed a method to find periods with low noise levels and perform the necessary recognition at these periods. If the signal contains such periods, it would be useful to use such methods together. 

In the field of image processing, Li et al. \cite{li2021wavecnet} proposed a network that integrates wavelet transforms into a CNN for denoising. 
This is also useful for extracting ECG features in conventional methods and it demonstrates excellent feature representation. We integrate a wavelet transform into an ECG denoising network and train the denoising parameters to improve the denoising performance of existing models.

\begin{figure}[htb]\centering
    \includegraphics[width=0.6\hsize]{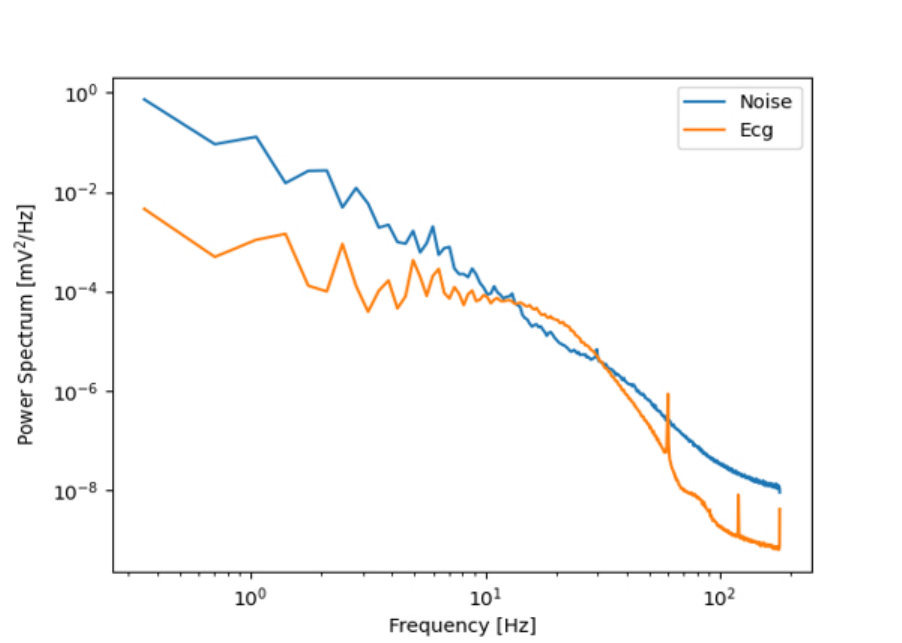}
	\caption{An example of power spectrum for an observed signal.}\label{Fig:ProblemSetting}
\end{figure}

The difficulty in denoising ECGs lies in the overlap between the noise and the ECG frequency bands, and when using a handmade threshold in a wavelet transform, there is a possibility of erasing valid information that should be kept. 
Fig. \ref{Fig:ProblemSetting} shows an example of the power spectra of ECG and noise for the data used in the experiment. 
There is overlap in all frequency bands of ECG and noise in the dataset we use in the experiment. In the ECG, there are frequency bands with large spike-like power, and by changing the filter strength for each frequency band, it is possible to remove the noise appropriately. In the range up to about 100 Hz, which is often the target of observation, the signal power in the low and high frequency bands is relatively small, so it may be possible to increase the filter strength in these frequency bands. 

\vspace{-0.1cm}
\section{Proposed Method}
\label{sec:proposed}

\subsection{Wavelet Layer}
\label{sec:proposed_wavelet_layer}

\begin{figure*}[b]
  \centering
    \includegraphics[width=0.8\hsize]{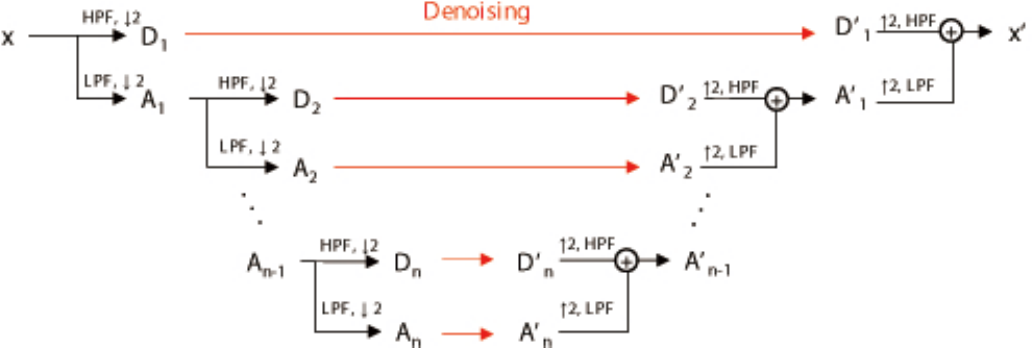}
	\caption{Summary of discrete wavelet transform. $D$ and $D^{\prime}$ indicate decomposed denoised component details. $D$ and $A$ are high- and low-frequency components, respectively.}\label{fig:summary_dwt}
\end{figure*}

An overview of the conventional discrete wavelet transform is shown in Fig. \ref{fig:summary_dwt}. 
As mentioned in Section \ref{sec:relatedwork}, wavelet transform denoising is designed for setting the filter after decomposing it into its components for each frequency band in order to give different filter strengths. 
First, the signal obtained by the HPF and LPF is down-sampled. Then, the HPF and LPF are applied to the output of the LPF among the down-sampled signals, and down-sampling is performed again. When repeated, the high-frequency component of each level is decomposed as Detail, and the low-frequency component of the last level is decomposed as Approximate. In conventional denoising, noise is removed by setting a threshold according to the frequency of each level and attenuating the component corresponding to noise. However, it is difficult to deal with complex or high-intensity noise with only a threshold value; additionally, the behavior of the filter must be changed according to the characteristics of the ECG. Therefore, we add a wavelet transform layer to the CNN and train different filters for each frequency to improve denoising performance.

\begin{figure*}[!b]\centering
    \includegraphics[width=0.70\hsize]{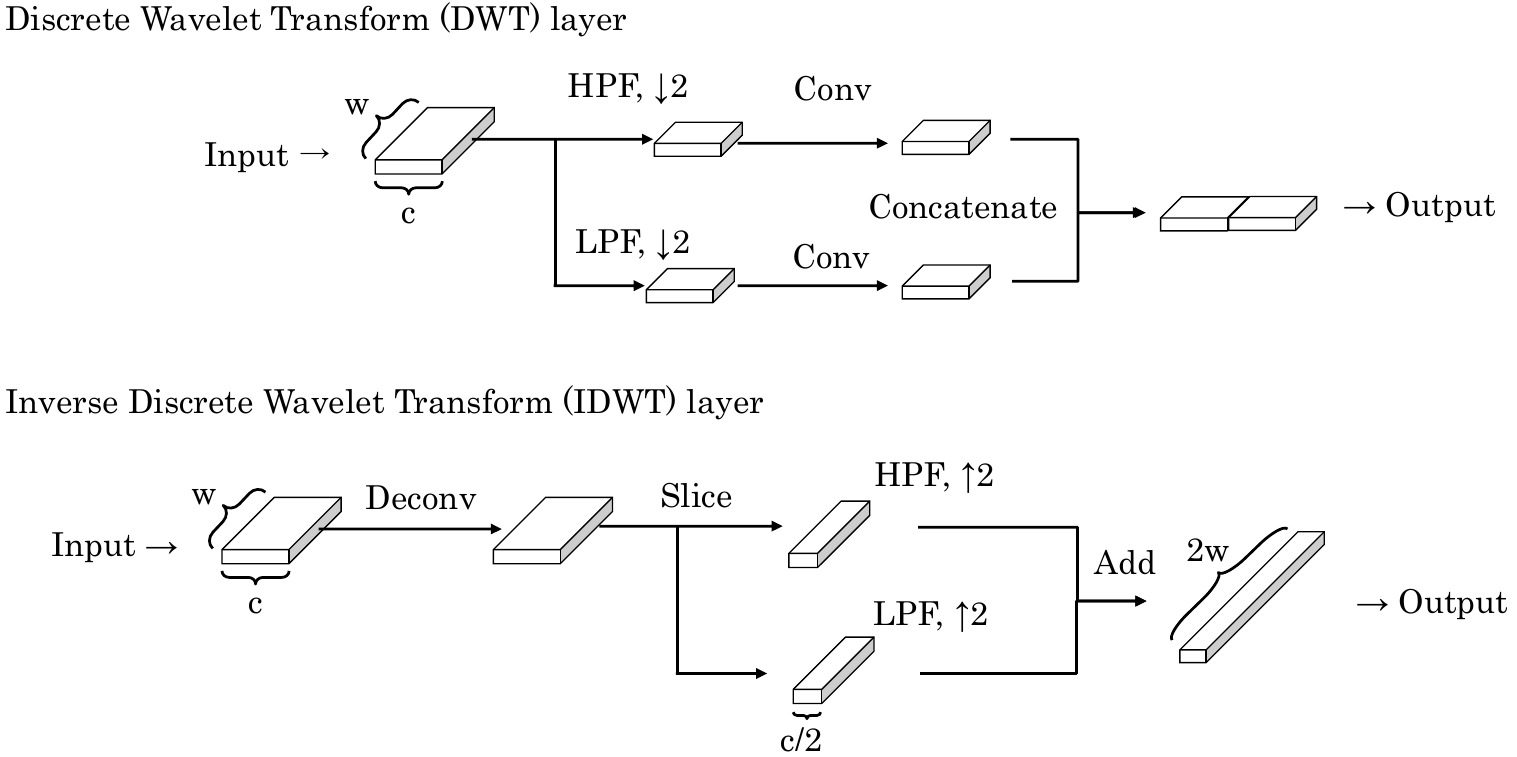}
	\caption{Discrete Wavelet Transform (DWT) and Inverse Discrete Wavelet Transform (IDWT) layers.}\label{fig2}
\end{figure*}
\begin{figure*}[bt]\centering
	\includegraphics[width=0.70\hsize]{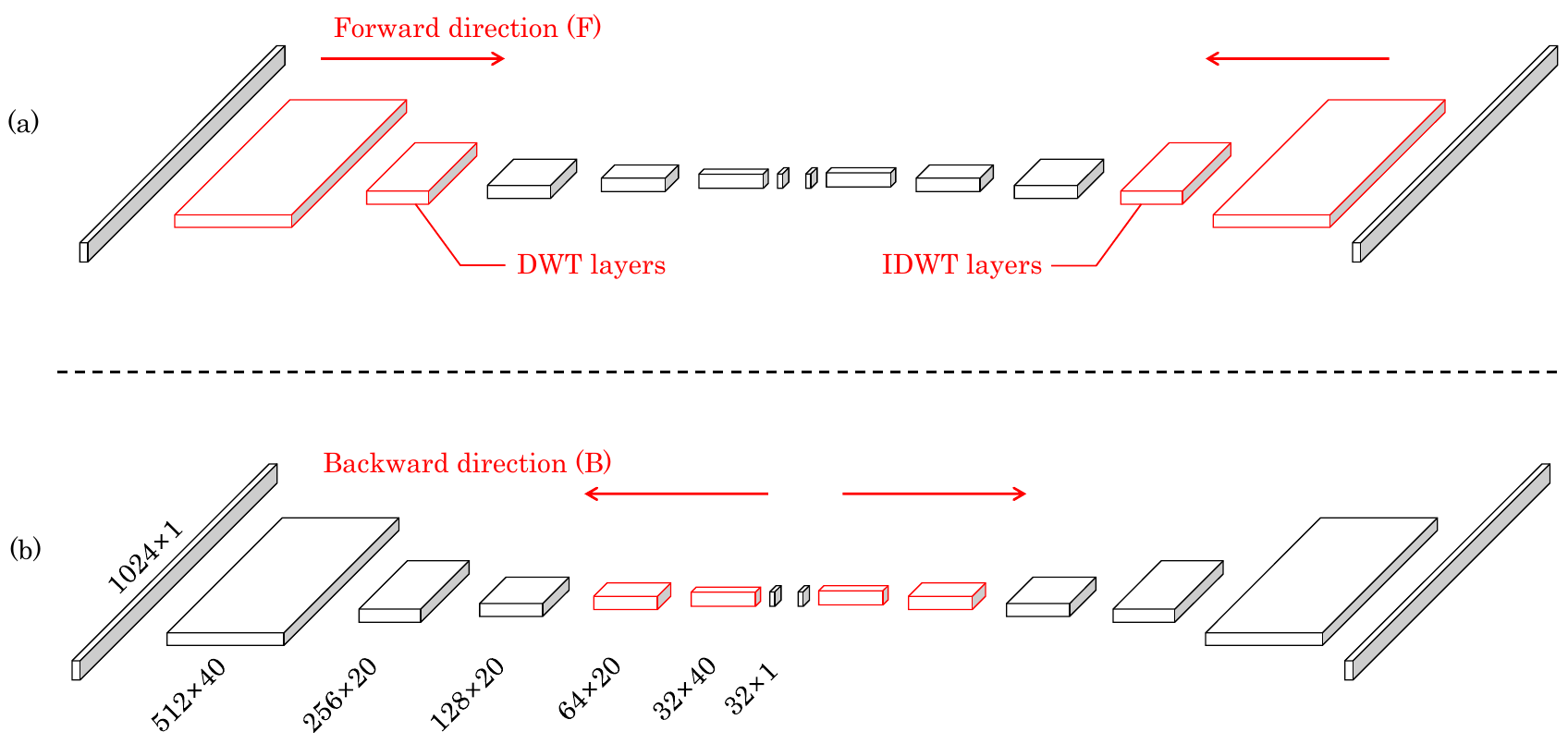}
	\caption{Wavelet integrated convolutional neural network: (a) Forward(F)-type; (b) Backward(B)-type introduction of DWT and IDWT layers. }\label{fig3}
\end{figure*}

Therefore, we introduced a CNN architecture to improve the denoising performance. 
For each level of the signal to be downsampled, the CNN can learn the optimal thresholds obtained from the training data. 
In addition, to make the network learn the behavior at each frequency level, we replace some convolutional layers in the encoder with Discrete Wavelet Transform (DWT) layers and some convolutional layers in the decoder with Inverse Discrete Wavelet Transform (IDWT) layers. By adding wavelet transform layers to the CNN and training different filters for each frequency level, we can expect to improve the denoising performance.

The proposed wavelet and inverse wavelet layers are shown in Fig. \ref{fig2}. The wavelet layer first performs a wavelet transform to decompose the input features into high- and low-frequency components. Then, convolutions are performed on each feature, and these are concatenated in the channel direction. 
By explicitly specifying the convolution layer weights for the HPF and LPF and optimizing them, we get different optimized filters for different frequency bands.

Meanwhile, the inverse wavelet layer performs the transpose convolution symmetrically with the wavelet layer of the encoder. Then, an HPF and an LPF are applied to each feature map for up-sampling, and each feature map is added to form the output. During training, we aim to improve the model’s robustness to complex noise by learning the parameters of the convolution layer for each frequency level.

The input and output dimensions are the same so that the convolutional layer, DWT, and IDWT are compatible. On the encoder side, the output dimension is half of the input dimension, and on the decoder side, the output dimension is twice the input dimension. It may be possible to further optimize these parameters, but we did not consider this part in this study to avoid making the discussion too complicated. This is an issue for future work.

\subsection{CNN model with wavelet layer}
\label{sec:proposed_cnn}

One possibility is to replace all convolutional layers with DWT and IDWT, but experiments have shown that this is not the best choice.  
In deep-learning models, it is known that deeper models better perform high feature extractions \cite{he2016deep}. However, if the model is separated by the frequency at a stage where the extraction of features is insufficient, accurate extraction ability may be reduced. In our experiment, we show that the ECG feature extraction capability can be improved without increasing the model parameters by retaining the CNN in its shallow layers and performing frequency separation using DWT and IDWT in the deep layers. To clarify the effect of DWT and IDWT layers, we prepared models with different numbers of layers and ablation studies are conducted. 

Fig. \ref{fig3}(a) shows a network that replaces some layers from the shallow layer with DWTs in the forward direction, and also replaces the corresponding IDWTs just before the output.
Fig. 3(b) shows the opposite model, where some layers from the deep layer are replaced by DWTs in the backward direction, and the corresponding IDWTs in the deep layer are also replaced just before the output. The replacement from the shallow layer is called forward (F), and the replacement from the deep layer is called backward (B). We also prepared models with all convolutional layers and all wavelet layers, 
which correspond to conventional networks. 
Through experiments, we will verify how many layers can be replaced with DWTs and IDWTs to obtain optimal noise denoising performance.

\section{Experimental Condition}
\label{sec:experimental_condition}
The dataset and network parameters used in the experiments are described in this section.

\subsection{Dataset}
\label{sec:experiment_dataset}
To clarify the effectiveness of the proposed method, the MIT-BIH arrhythmia dataset \cite{moody2001impact} and the MIT-BIH noise stress test dataset (MIT-NST) \cite{moody1984noise} were used as the ECG dataset and the noise dataset, respectively. The MIT-BIH dataset contains ECG data from 48 patients, each with a sampling frequency of 360 Hz and a measurement time of 30 min.The MIT-NST contains BW, EM, and MA noise, and all were used for evaluation. In the experiments, the first 90\% of the data was used for training, and the remainder was for evaluation. To preserve the most accurate model, the last 20\% of the training data was used for validation. For training, patients 102 and 104 were omitted because they included paced peaks. To apply the CNN model, 3 s of data was extracted with a window width of 1,024, as in the conventional method. For each patient, we randomly selected 160 samples for training and 40 for validation for a total of 200 samples. There were 9,200 total training samples, including validation.

The MIT-BIH dataset is band-pass filtered in the range of 0.1---100 Hz \cite{moody2001impact}, but because the correct data also contain noise, a fifth-order HPF with a cutoff frequency of 0.67, an LPF at 100 Hz, and a smoothing filter with a kernel size of five were applied before dividing the data by window size. After dividing the data into windows, power per window was calculated to exclude outliers, and only samples within the upper 95th and lower 5th percentiles were used. As the amplitude of the signal differed for each patient, normalization was performed for each patient. The SNRs for the training and validation data were -2.5, 0, 2.5, 5.0, and 7.5, and those for evaluation were -10, -7.0, -3.0, -1.0, 3.0, 7, and 10. 
In order to verify whether the system can handle noise at levels not used in training, the test was conducted using a wider range of noise than in training. 

\subsection{Model Parameters}
\label{sec:experiment_model_parameters}
In this experiment, the denoising performance of the proposed wavelet layer was evaluated. The fully convolutional network model \cite{chiang2019noise} was used as the conventional model. This model is an autoencoder that uses only convolution operations, making it possible to compress features with lengths of 1,024 to 32.

The parameters in each layer are shown in Table I. The stride of the convolution of each layer was two, and the dimension was compressed during convolution. The number of feature maps was set to 40 for the first and second-to-last convolutional layers and 20 for the other layers to reduce the number of parameters during training while increasing accuracy by extracting more input and output features. The parameters of a window width of 1,024, as in the conventional method. For the decoder were in reverse order of the encoder. There was no contraction path between the encoder and decoder; thus, it can also be employed in systems where the encoder and decoder are placed in different locations to perform compression and reconstruction. To apply the wavelet transform in the wavelet layer, the wavelet Daubechies 6 (db6) coefficient was used as the mother wavelet. This coefficient is similar to an ECG morphology; thus, it was used in the conventional methods \cite{banerjee2012delineation,jenkal2016efficient}.
	
For the convolutional and wavelet layers, apart from the last outputs of the encoder and decoder, a batch normalization layer, an exponential linear unit for the activation function, and a dropout layer with 10\% probability were applied. An Adam optimizer with a learning rate of 0.0001 was used to optimize training. The batch size was 200, and 200 epochs were trained. Training and evaluation were performed 10 times, and the average value was used as the result. The implementation was performed in Python v.3.8 and TensorFlow v.2.4.

\section{Experimental Results}
\label{sec:experimental_result}

\subsection{Quantitative Evaluation}
\label{sec:experimental_quantitative}

\begin{table}[!b]
	\caption{Parameters of training model. The numbers in Conv and Deconv represent the number of filters, kernel size, and stride, in order. This represents a backward-type model comprising three wavelet layers.}\label{table:param_model}
	\begin{center} \scriptsize
	\begin{tabular}{ccccc}
		\hline
		& No & Output & FCN \cite{chiang2019noise} & Proposed \\
		\hline \hline
		Input & - & 1024 $\times$ 1 & & \\
		\hline
		Encoder & 1 & 512 $\times$ 40 & Conv(40, 16, 2) & Conv(40, 16, 2) \\
			& 2 & 256 $\times$ 20 & Conv(20, 16, 2) & Conv(20, 16, 2) \\ 
			& 3 & 128 $\times$ 20 & Conv(20, 16, 2) & HPF, Conv(20, 8, 2) \\
			& & & & LPF, Conv(20, 8, 2) \\ 
			& 4 & 64 $\times$ 20 & Conv(20, 16, 2) & HPF, Conv(20, 8, 2) \\
			& & & & LPF, Conv(20, 8, 2) \\ 
			& 5 & 32 $\times$ 40 & Conv(20, 16, 2) & HPF, Conv(20, 8, 2) \\
			& & & & LPF, Conv(20, 8, 2) \\ 
			& 6 & 32 $\times$ 1 & Conv(1, 16, 1) &  Conv(1, 16, 1) \\
		\hline
		Decoder & 7 & 32 $\times$ 1 & Conv(1, 16, 1) &  Conv(1, 16, 1) \\
			& 8 & 64 $\times$ 40 & Conv(40, 16, 2) & Conv(40, 16, 2) \\
			& 9 & 128 $\times$ 20 & Conv(20, 16, 2) & Conv(20, 16, 2) \\
			& 10 & 256 $\times$ 20 & Conv(20, 16, 2) & Conv(20, 16, 2) \\
			& 11 & 512 $\times$ 20 & Conv(20, 16, 2) & Conv(20, 16, 2) \\
			& 12 & 1024 $\times$ 40 & Conv(40, 16, 2) & Conv(40, 16, 2) \\
			\hline
		Output & 13 & 1024 $\times$ 1 & Conv(1, 16, 1) & Conv(1, 16, 1) \\
		\hline
	\end{tabular}
    \end{center}
\end{table}

\begin{table*}[!b]
	\caption{Calculation results for RMSE. B and F are backward- and forward-types (Fig. \ref{fig3}), respectively. FCN: fully convolutional network; RMSE: root mean-square error; SNR: signal-to-noise ratio.}\label{table:result_rmse}
		\centering \scriptsize
		\begin{tabular}{cccc|ccccccc}
			\hline
			 & & & & \multicolumn{7}{c}{Input SNR} \\
			ID & Model & Wavelet & Type & -10 & -7 & -3 & -1 & 3 & 7 & 10 \\
			\hline \hline
			1 & FCN \cite{chiang2019noise} & - & - & 0.2104 & 0.1580 & 0.1141 & 0.0997 & 0.0806& 0.0707 & 0.0670 \\ \hline
			2 & & 1 & F & 0.2094 & 0.1575 & 0.1140 & 0.0996 & 0.0803 & 0.0701 & 0.0664 \\ 
			3 & & 2 & F & 0.2011 & 0.1533 & 0.1119 & 0.0977 & 0.0785 & 0.0686 & 0.0651 \\
			4 & & 3 & F & 0.1935 & 0.1489 & 0.1090 & 0.0953 & 0.0771 & 0.0679 & 0.0646 \\
			5 & & 4 & F & 0.1814 & 0.1423 & 0.1063 & 0.0936 & 0.0766 & 0.0683 & 0.0654 \\
			6 & Proposed & 1 & B & 0.1811 & 0.1399 & \textbf{0.1031} & \textbf{0.0907} & \textbf{0.0743} & \textbf{0.0663} & \textbf{0.0636} \\
			7 & & 2 & B & 0.1817 & 0.1403 & 0.1036 & 0.0911 & 0.0749 & 0.0670 & 0.0644 \\
			8 & & 3 & B & 0.1774 & 0.1393 & 0.1039 & 0.0915 & 0.0752 & 0.0673 & 0.0648 \\
			9 & & 4 & B & \textbf{0.1772} & \textbf{0.1392} & 0.1045 & 0.0922 & 0.0761 & 0.0683 & 0.0657 \\
			10 & & 5 & - & 0.1822 & 0.1417 & 0.1049 & 0.0922 & 0.0758 & 0.0680 & 0.0655 \\
			\hline
		\end{tabular}
		\\ 
		\centering
\end{table*}

\begin{table*}[!t]
	\caption{Calculation results for SNR improvement. B and F are backward- and forward-types (Fig. \ref{fig3}), respectively. FCN: fully convolutional network; SNR: signal-to-noise ratio.}\label{table:result_snrimp}
		\centering \scriptsize
		\begin{tabular}{cccc|ccccccc}
			\hline
			 & & & & \multicolumn{7}{c}{Input SNR} \\
			ID & Model & Wavelet & Type & -10 & -7 & -3 & -1 & 3 & 7 & 10 \\
			\hline \hline
			1 & FCN \cite{chiang2019noise} & - & - & 18.51 & 18.10 & 17.08 & 16.30 & 14.14 & 11.26 & 8.72 \\ \hline
			2 & & 1 & F & 18.55 & 18.13 & 17.10 & 16.32 & 14.19 & 11.33 & 8.80 \\
			3 & & 2 & F & 18.89 & 18.36 & 17.26 & 16.49 & 14.40 & 11.54 & 8.99 \\
			4 & & 3 & F & 19.20 & 18.62 & 17.52 & 16.73 & 14.57 & 11.66 & 9.08 \\
			5 & & 4 & F & 19.77 & 19.04 & 17.76 & 16.91 & 14.64 & 11.63 & 8.99 \\
			6 & Proposed & 1 & B & 19.82 & 19.23 & \textbf{18.07} & \textbf{17.23} & \textbf{14.95} & \textbf{11.92} & \textbf{9.28} \\ 
			7 & & 2 & B & 19.80 & 19.20 & 18.03 & 17.19 & 14.89 & 11.83 & 9.17 \\
			8 & & 3 & B & 19.96 & \textbf{19.24} & 17.99 & 17.14 & 14.84 & 11.79 & 9.13 \\
			9 & & 4 & B & \textbf{19.97} & \textbf{19.24} & 17.94 & 17.07 & 14.72 & 11.65 & 8.99 \\
			10 & & 5 & B &  19.74 & 19.09 & 17.90 & 17.06 & 14.77 & 11.70 & 9.04 \\
			
			\hline
		\end{tabular}
\end{table*}

\begin{figure*}[t]\centering
	\begin{tabular}{cc}
		\begin{minipage}{0.48\linewidth}
		\includegraphics[width=0.98\hsize]{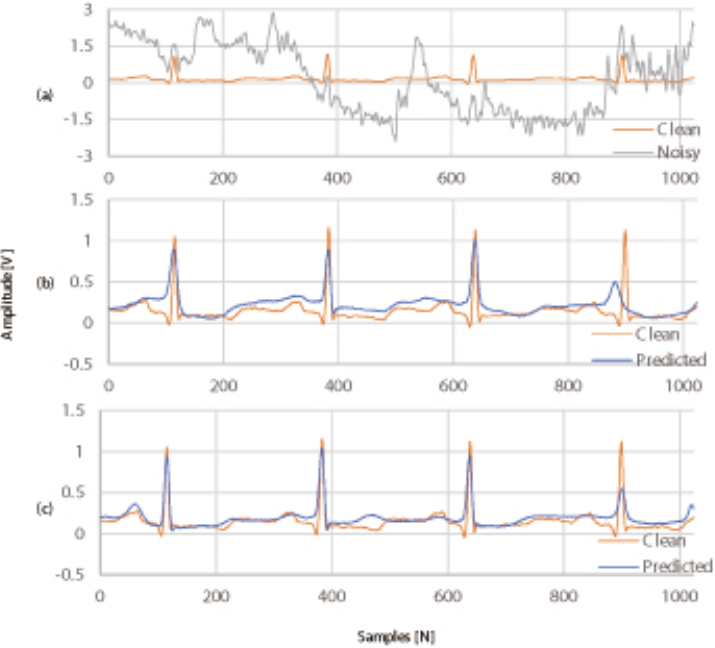}
		\caption{Outputs of the convolutional neural network when patient is 100 and signal-to-noise ratio is -10: (a) Noisy electrocardiogram; (b) Fully convolutional network (FCN); (c) Backward-type discrete wavelet transform (DWT) FCN with one DWT layer.}\label{fig4}
		\end{minipage}
		\hspace{2mm}

		\begin{minipage}{0.48\linewidth}
		\includegraphics[width=0.98\hsize]{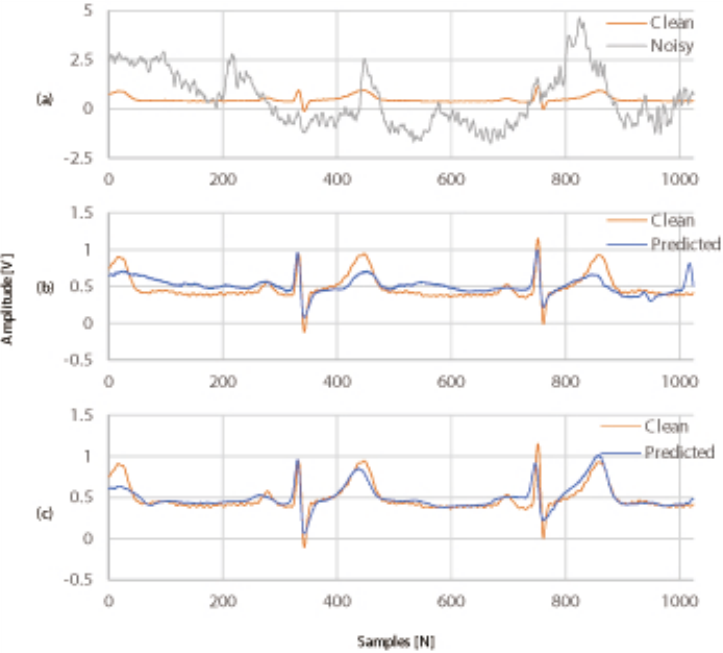}
		\caption{Outputs of convolutional neural network when patient is 117 and signal-to-noise ratio is -3; (a) Noisy electrocardiogram; (b) Fully convolutional network (FCN); (c) Backward-type discrete wavelet transform (DWT) FCN with one DWT layer.}\label{fig5}
		\end{minipage}
	\end{tabular}
\end{figure*}

For evaluation metrics, root mean squared error ($RMSE$) and SNR improvement ($SNR_{imp}$) were applied.  
The experimental results of RMSE and SNR improvement are shown in Tables \ref{table:result_rmse} and \ref{table:result_snrimp}, respectively. 
A smaller value is better for RSME, and a smaller value is better for SNR improvement. 
Each ID in the tables shows different experimental conditions. ID1 represents the conventional CNN model, ID2--ID5 represent the proposed model that increases the wavelet layer in the forward direction (Fig. \ref{fig3}(a)), ID6--ID9 represent the proposed model that increases the wavelet layer in the backward direction (Fig. \ref{fig3}(b)), and ID10 shows the model with all wavelet layers. These different models are forward-, backward-, and all-wavelet-layer-type fully convolutional networks (FCNs), respectively.

In Table \ref{table:result_rmse}, ID6 shows the best RMSE when SNRs are in the range of -3--10, and ID9 shows the best RMSE when SNRs are -10 and -7. ID6 and ID9 are backward-type models with one wavelet layer and three wavelet layers, respectively. The RMSE differences between the FCN and the best models are 0.0293 and 0.0034 when SNRs are -10 and 10. The backward-type model shows the best RMSE in all SNRs, and there is more improvement when high-intensity noise is infected.

When focusing on the order of wavelet layers, there is no significant difference between the FCN and ID2. In the case of the forward-type model, the RMSE improves when the wavelet layer increases, as shown in ID2--ID5. However, the RMSE of ID5 is not better than that of ID6, which is a backward-type model. Therefore, the performance of the forward-type model is inferior to that of the backward-type model. Furthermore, in the case of all wavelet models, the RMSE differences between FCN and ID10 are 0.0282 and 0.0015, respectively, when SNRs are -10 and 10. The performance of all wavelet models is better than that of the FCN, but it is inferior to the backward-type model. From these results, it is confirmed that the backward-type model shows the best performance.

As shown in Table \ref{table:result_snrimp}, SNR improvements show similar trends to the RMSE. Specifically, ID6 is the best when the SNRs are in the range of -3--10, and ID9 shows the best when the SNRs are -10 and -7. 
When comparing the PRD of ID1 and ID6 with and without the wavelet layer, the PRDs improve by 10.19 and 1.04 when the SNR is -10 and 10. 
On the other hand, the SNR improvement of ID1 and ID6 improves 1.32 and 0.56, respectively, when the SNR is -10 and 10. Because the best model is the same as the RMSE, it can be concluded that the backward-type model improves denoising performance.

\subsection{Qualitative Evaluation}
\label{sec:experimental_qualiative}

For qualitative evaluation, the output results of samples 100 and 117 as different ECG waveforms are shown in Figs. \ref{fig4} and \ref{fig5}, respectively. In Fig. \ref{fig4}, the proposed method extracts the signal more faithfully, and the shape before the first R-wave is closer to the clean signal. In the conventional method, the QRS waveform is distorted due to low-frequency noise in the second and third R-waves. In addition, the peak of the fourth R-wave is misaligned, whereas the peak of the proposed method is correct.

In Fig. \ref{fig5}, there is no difference in the shape of the R-wave between the two methods, but the ST-wave waveform of the proposed method is more like the original. Furthermore, in this sample, the conventional method incorrectly recovers a peak that does not exist in the proposed method. On the other hand, the output of the proposed method does not show such a peak, indicating that the proposed method is superior to the original method. 

\begin{figure*}[!t]
	\centering
	\includegraphics[width=0.95\hsize]{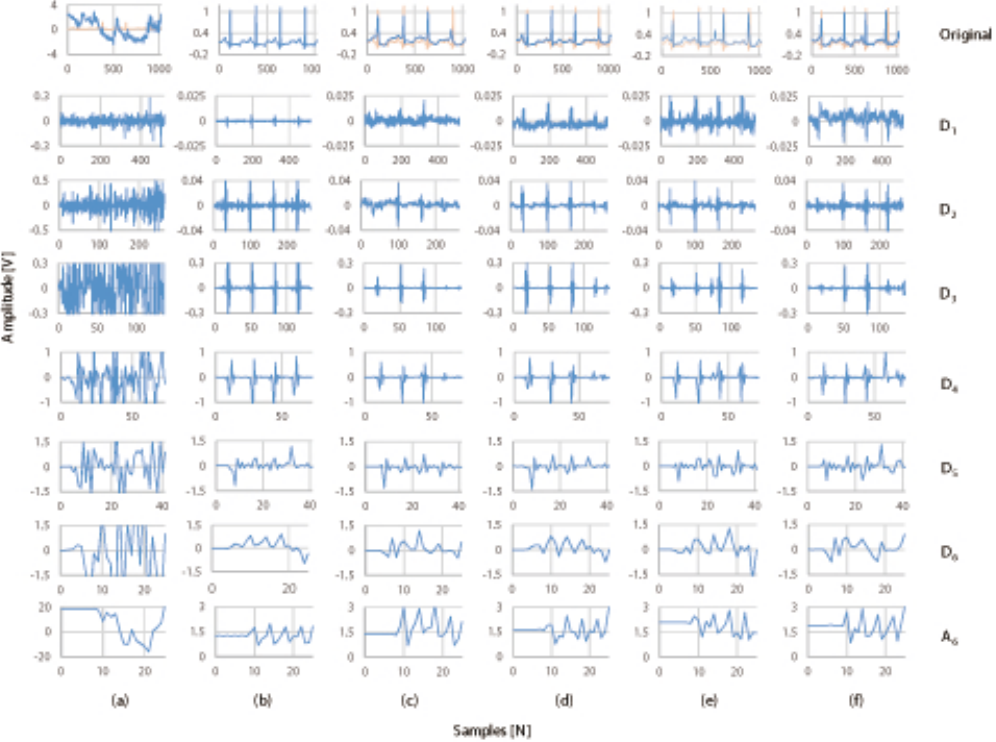}
	\caption{Result of discrete wavelet transform (DWT) decomposition when patient is 100 and signal-to-noise ratio is -10: (a) Noisy electrocardiogram (ECG); (b) Clean ECG; (c) Output of fully convolutional network (FCN) for ID1; (d) Backward DWT FCN (ID6); (e) Forward DWT FCN (ID2); (f) All DWT-layer-type FCN (ID10).}\label{fig6}
\end{figure*}

Furthermore, the wavelet transform is applied to the reconstruction results of ID1, ID2, ID6, and ID10 to clarify which frequency bands are recovered well, depending on the type of wavelet layer. The results of the wavelet transform are shown in Fig. \ref{fig6}. When comparing the results of ID1 and ID6, there is a difference in D3, which is the frequency band between 22.5 and 45 Hz. For example, the original ECG has four peaks, but the output of ID1 does not have a peak at the fourth location. ID2 and ID10, which are less accurate than ID6, differ in D1 from 45--90 Hz, indicating that they contain many frequency components that are not present in the original ECG signal. Therefore, it can be concluded that the noise of the high-frequency components is not removed, including ID1 and ID6.

From these results, it is confirmed that the wavelet layer improves the ECG denoising performance on the CNN, both quantitatively and qualitatively.

\section{Discussion}
\label{sec:discussion}
In our experiments, a comparison of all convolutional layers, a forward-type model, a backward-type model, and an all-wavelet-layers model was performed. It was shown that the backward-type model with one wavelet layer had the best accuracy and robustness to noise.

Although the wavelet layer improved the denoising performance of the conventional CNN, the accuracy of models with more wavelets in the forward direction and those with all wavelet layers was degraded. This is thought to be related to the frequency range of the ECG. Many theories suggest that an ECG contains important features between 0.5 and 40 Hz. The first layer of our model represents the components in the range of 90--180 Hz in the sampling frequency. When we separated the low- and high-frequency components in the shallow layer, the noise-removal capability of the high-frequency components was insufficient, and the accuracy was reduced. Thus, as the layers of the CNN became deeper in the feature extraction stage, they became closer to the frequency band that originally represented the ECG. Thus, replacing the deeper layers with wavelet layers may have led to improved accuracy. Additionally, the proposed model did not use a contraction path, which is used in existing methods to improve accuracy \cite{singh2020new}; hence, recovery accuracy is insufficient and needs further improvement. A related future challenge is to modify the network structure and parameters to create a model with higher accuracy, even under severe hardware constraints, such as at edge terminals.

\section{Conclusion}
\label{sec:conclusion}
This paper proposed a model with a wavelet layer for ECG denoising that is more robust to noise. The wavelet layer separates high and low frequencies, which is expected to improve the denoising performance for frequency bands where noise and ECG overlap.

In the experiments, the different model types were compared on different SNRs from -10 to 10, and the model with one wavelet layer in the backward direction showed the highest accuracy compared with the conventional model. The improvement was observed qualitatively and quantitatively; thus, it was confirmed that the wavelet layer was effective for ECG denoising. In the future, we aim to propose a model that is lighter and has higher denoising accuracy.

\begin{credits}
\subsubsection{\ackname} This work was supported by JST PRESTO Grant Number JPMJPR2135 and JSPS KAKENHI Grant Number JP20H04472. 
\end{credits}

\bibliographystyle{splncs04}

\end{document}